\documentclass[onecolumn,showkeys,preprintnumbers,aps,a4paper,amssymb,prd,superscriptaddress,nofootinbib]{revtex4-2}
\usepackage{comment}
\usepackage{graphicx}
\usepackage{epsf}
\usepackage{bm}
\usepackage{amsmath}
\usepackage{amsfonts}
\usepackage{amssymb}
\usepackage{epstopdf}
\usepackage{color}
\usepackage[dvipsnames]{xcolor}
\usepackage{verbatim}
\usepackage{multirow}
\usepackage{soul}
\usepackage{physics}
\usepackage{bm}

\usepackage[width=0.00cm, height=0.00cm, left=1.50cm, right=1.50cm, top=2.00cm, bottom=2.00cm]{geometry}
\usepackage{microtype}
\usepackage[colorlinks = true,
            linkcolor = teal,
            urlcolor  = black,
            citecolor = teal,
            anchorcolor = blue]{hyperref}
\usepackage[capitalize]{cleveref}
\usepackage[normalem]{ulem}
\usepackage{enumitem}
\usepackage{booktabs}

\usepackage{lipsum}

\makeatletter\let\expandableinput\@@input\makeatother

\begin{document}
%title{Exploring $\Lambda_s $CDM model with new DESI BAO 2024}

%\title{Consistency of Sign-switching cosmological constant model with DESI BAO 2024}
\begin{center}
		\vspace{0.4cm} {\large{\bf \textbf{Beyond $\Lambda$CDM: Exploring a Dynamical Cosmological Constant Framework Consistent with Late-Time Observations}}} \\
		\vspace{0.4cm}
		
		\normalsize{ Archana Dixit$^1$, Manish Yadav$^2$, Anirudh Pradhan $^3$, M. S. Barak$^4$ }\\
		\vspace{5mm}	
		\normalsize{$^{1 }$ Department of Mathematics, Gurugram University Gurugram, Harayana, India.}\\
		\normalsize{$^{2,4}$ Department of Mathematics, Indira Gandhi University, Meerpur, Haryana 122502, India.}\\
        	\normalsize{$^{3 }$ Centre for Cosmology, Astrophysics and Space Science (CCASS), GLA University, Mathura, Uttar Pradesh, India.}\\ 
		\vspace{2mm}
		    $^1$Email address: archana.ibs.maths@gmail.com\\
			$^2$Email address: manish.math.rs@igu.ac.in\\
                $^3$Email address: pradhan.anirudh@gmail.com\\
             $^4$Email address: ms$_{-}$barak@igu.ac.in\\
\end{center}

{\bf Abstract:} 
In this work, we investigate a cosmological scenario with a time-dependent cosmological constant $\Lambda$(t) within the spatially flat Friedmann–Lemaître–Robertson–Walker (FLRW) framework. Here we study a power-law $\Lambda(t)$CDM model characterized by a dynamic cosmological constant expressed as a function of the Hubble parameter and its derivative $\Lambda(t)$ $=\alpha (\dot H+H^{2})+\lambda H^2+4\pi G\rho\eta.$ Using recent observational datasets (DESI BAO, OHD, and PP\&SH0ES), we constrain the model’s free parameters $(H_{0},\alpha,\lambda,\eta)$ and analyze their impact on key cosmological quantities. A Markov chain Monte Carlo (MCMC) analysis of the best-fit value of $H_{0}=71.9\pm 0.23$ km/s/Mpc from PP\&SH0ES analysis only, which substantially alleviates the existing tension between early and late-time determinations of the Hubble constant, reducing it to $\sim1.5\sigma$. The reconstructed $Om$ diagnostic exhibits a negative slope, indicating a dynamic dark energy behavior with quintessence-like characteristics ($\omega>-1$).  These results suggest that the proposed $\Lambda(t)$ model provides a viable alternative to the standard $\Lambda$CDM paradigm to explain the late-time acceleration of the universe. Our findings show that this model alleviates the Hubble tension more effectively than the standard $\Lambda$CDM . The model also demonstrates compatibility with late-time Hubble parameter observations and offers a compelling framework to address the limitations of $\Lambda$CDM.\\

\section{Introduction}

The first evidence for the universe is expanding at an accelerating rate by two research groups, the High-Z Supernova, and Supernova Cosmology Project Teams, after that, this has been confirmed by the astrophysical observation data such as  Type Ia Supernovae (SNe Ia) \cite{ref1,ref2,ref3}, the Cosmic Microwave Background Radiation (CMB) \cite{ref4,ref5,ref6}, Baryonic Acoustic Oscillations (BAO)\cite{ref7,ref8,ref9,ref10},  and precise determinations  value of the Hubble constant \cite{ref11,ref12,ref13,ref14}. Meanwhile, in modern cosmology, the phenomenon of an accelerated universe is a directed correlated unknown foam of energy known as dark energy, which exhibits negative pressure and a dominant component in the total energy budget of the universe \cite{ref15,ref16,ref17}. Although the influence of dark energy on the universe accelerated expansion is widely recognized, the fundamental nature of dark energy remains one of the most profound unresolved challenges in theoretical physics. \\

 Over the last few decades, the scientific community has embraced $\Lambda$CDM as the standard model in cosmology based on the theory of inflation \cite{ref18,ref19,ref20,ref21} . This model has excellent agreement with a wide range of observational data from various regions of the universe \cite{ref22,ref23,ref24,ref25}. Nevertheless, recent studies have detected certain inconsistencies have emerged in cosmological theory, one  of them a significant matter of concern fine-tuning problem \cite{ref26}. Observations reveal that the cosmological constant, denoted as \( \Lambda \), possesses an exceedingly small value, enough to explain the observed acceleration of the cosmos. However, predictions made through quantum field theory indicate a vastly larger value for \( \Lambda \), with discrepancies that arise from a staggering factor of \( 10^{120} \). Additionally, the coincidence problem \cite{ref27} highlights the puzzlingly similar orders of magnitude between \( \Lambda \) and the density of ordinary baryonic matter within the universe. These problems motivate researchers to explore frameworks beyond the standard $\Lambda$CDM model and to consider new physics that may help resolve these issues. In the investigation of the accelerated  universe, researchers have mainly adopted two primary methodologies to engage with various cosmological models. The first approach involves modifications to classical general relativity (GR) \cite{ref27a,ref27b,ref27c,ref27d,ref27e,ref27f,ref27g,ref27h}, and the second approach introduces enigmatic energy components, widely referred to as dark energy (DE). These challenges have prompted the search for alternative models that offer time-dependent equations of state to account for the accelerating expansion. Some of these alternatives include Chaplygin gas \cite{ref28,ref29,ref29a,ref29b,ref29c}, tachyon fields \cite{ref30,ref30a,ref30b,ref30c},  K-essence \cite{ref31,ref32,ref32a,ref32b}, quintessence \cite{ref33,ref33a,ref33b,ref33c}, and phantom energy \cite{ref34,ref34a,ref34b,ref35}, which propose different mechanisms for the observed acceleration.\\

In the 20th century, various kinds of models have been suggested in the cosmology field that involve time-dependent modifications
to the cosmological constant/vacuum energy, $\Lambda(t)$, or considering the decay of the vacuum, often introducing a specific cosmic time \cite{ref36,ref37,ref38}. These models explore various phenomenological decay behaviours for $\Lambda(t)$, derived either through quantum mechanical arguments \cite{ref39} or spacetime geometric approaches \cite{ref40,ref41}. Researchers have also investigated the interactions between matter density and vacuum energy through different methodologies, comparing these models with the most recent cosmological data \cite{ref42,ref43,ref44}. One potential solution to the shortcomings of the $\Lambda(t)$CDM model ( where $\Lambda(t)$ is the time-varying cosmological constant foam of DE, and CDM is the cold dark matter  is the Running Vacuum Model (RVM)). This theoretical framework, which is anchored in quantum field theory within a curved spacetime context, characterizes the vacuum energy density as a power series expansion comprising the Hubble parameter and its derivatives with respect to cosmic time. This model has demonstrated notable success in capturing cosmological observations and, in some cases, has provided a better fit than the $\Lambda$CDM model \cite{ref45,ref46,ref47}. For a deeper understanding of the RVM and insights into recent theoretical developments in the field, readers are encouraged to consult the relevant references~\cite{ref48,ref49}. Overduin and Cooperstock~\cite{ref38} examined how the evolution of the scale factor is influenced by a varying cosmological term, considering forms such as $\Lambda = A t^{-l}$, $\Lambda = B a^{-m}$, $\Lambda = C t^{n}$, and $\Lambda = D q^{r}$, in their analysis, the coefficients $A$, $B$, $C$, $D$, and the exponents $l$, $m$, $n$, $r$ are treated as constants. In this work, we specifically adopt a power-law dependence for the cosmological constant term based on the cosmic time, expressed as $\Lambda(t) = \alpha(\dot{H}+H^2) + \lambda H^2 + 4\pi G \rho \eta$,  where $\alpha$, $\lambda$, and $\eta$ are arbitrary constants, and $G$ and $\rho$ denote the gravitational constant and the total energy density of the universe, respectively. A similar approach was adopted in ref.\cite{ref50}, the author used a phenomenological approach to describe the time evolution of $\Lambda(t)$, representing it through a linear combination involving the Hubble parameter and its derivatives. Recently, several important studies on $\Lambda(t)$CDM models employing the power-law form with various modifications have been reported in the literature \cite{ref51,ref52,ref53,ref54}.
 
In the standard cosmological framework, the cosmological constant 
$\Lambda$ is assumed to be constant; however, such a fixed term faces major theoretical challenges, such as the fine-tuning and cosmic coincidence problems. To address these issues, several studies have suggested allowing $\Lambda(t)$ to vary with cosmic time or with the Hubble parameter\cite{ref54a,ref54b,ref54c}, which can provide a more natural explanation for the observed late-time accelerated expansion of the Universe. Motivated by this, we reformulate the model by introducing a generalized parameterization of $\Lambda(t)$  as a function of $\dot H$ $H^{2}$, and $\rho$. This approach offers additional flexibility to capture possible deviations from the standard $\Lambda$CDM  model, while remaining consistent with the Friedmann equations. Such a formulation enables us to probe potential dynamical behavior of vacuum energy and to test whether observational data (DESI BAO, OHD, and PP\&SH0ES) favor any evolution in $\Lambda(t)$ over cosmic time, which would provide new phenomenological insights into the nature of dark energy\cite{ref54d,ref54e,ref54f} .
The present manuscript is organized into several key sections. In section II, we investigate the FLRW universe within the paradigm of time-varying cosmological constant $\Lambda$(t) and create a power law model based on $\Lambda$(t) in terms of the Hubble parameter. In the next section III, we introduce the latest DESI BAO, OHD, and PP\&SH0ES data sets and the methodology employed in our analysis. Section IV  presents the results of the analysis and discusses the key findings. Lastly, Section V offers a final summary and conclusion.

\section{MODEL}\label{sec2}

First, we consider the FLRW metric, which can be formulated in cartesian coordinates system (t,x,y,z) as follows:
\begin{equation}
\label{matrix}
\text{d}s^2=-\text{d}t^2 + s(t)^2 \dfrac{\text{d}x^2 + \text{d}y^2 + \text{d}z^2}{\left[1 + \frac{\kappa}{4}(x^2 + y^2 + z^2)\right]^2}
%\text{d}s^2=-\text{d}t^2 + s^2\left[ \dfrac{\text{d}r^2}{1-\kappa r^2} + r^2(\sin{^2}\phi \text{d}\theta^2 + \text{d}\phi^2)\right] ,
\end{equation}
where $\kappa$ represents the curvature scalar, which represents open, flat, and closed universes with $\kappa < 0$, $\kappa = 0$ and $\kappa > 0$, respectively.The function 
s(t), referred to as the scale factor, evolves with cosmic time t. Also, the Einstein's field equations from General Relativity, are given by
 
 \begin{equation}\label{fieldeqn}  
 	G_{\mu\nu}\equiv R_{\mu\nu}-\frac{1}{2}g_{\mu\nu} R = 8\pi GT_{\mu\nu},
 \end{equation}
where $G_{\mu\nu}$, $R_{\mu\nu}$, $R$, and $g_{\mu\nu}$ are representing Einstein tensor, Ricci tensor, Ricci scalar, and metric tensor, respectively. Again, on the right hand side, $G$, and $T_{\mu\nu}$ are representing Newton’s gravitational constant and energy-momentum tensor, respectively. The most commonly used form of the total energy-momentum tensor  for a perfect fluid is:
 
\begin{equation}\label{emt}
 	T^{\nu}_{\,\,\mu} = \text{diag} [-\rho, p, p, p].    
 \end{equation}

As a consequence of the twice-contracted Bianchi identity, that is, $G^{\mu\nu}_{\;\;\;;\nu}=0$, the Einstein's field equations \eqref{fieldeqn} satisfy the conservation equation which is as follows: 

\begin{equation}\label{ece}
 	T^{\mu\nu}_{\;\;\;;\nu}=0.    
 \end{equation}
 Further, for a perfect fluid matter distribution, \eqref{ece} simplifies to
 \begin{equation}\label{ce}
 	\dot{\rho}+3H(1+ w)\rho=0,    
 \end{equation}
where the overdot at the top of any parameter represents its derivative with respect to
the cosmic time $t$. For a perfect fluid $i$, with pressure $p_i$, energy density $\rho_i$, and constant equation of state (EoS) defined by $w_i=p_i/\rho_i$, the continuity equation \eqref{ce} describes the evolution of its energy density  as follows:

 \begin{equation}\label{eos}
 	\rho_i=\rho_{i0}s^{-3(1+w_i)},    
 \end{equation}
 where $\rho_{i0}$ represents the present-day value of $\rho_{i}$, corresponding to $s_0=1$, the present-day value of the scale factor. The subscript "0" attached to any quantity denotes its value in the present-day Universe. We assume that the presence of standard cosmological components in the Universe, such as radiation (consisting of photons and neutrinos) characterized by an equation of state (EoS) parameter $w_{\rm r}=p_{\rm r}/\rho_{\rm r}=\frac{1}{3}$, pressureless fluid (including baryonic and cold dark matter) with EoS $w_{\rm m}=p_{\rm m}/\rho_{\rm m}=0$, and EoS of dark energy denoted as $w_{de}$ to be free by observations in the analysis. These energy constituents interact only through gravity. As a consequence, each energy source independently  satisfies the continuity equation \eqref{ce}, and considering \eqref{eos}, this leads to
 \begin{equation}\label{sources}
 	\rho\equiv\rho_\text{r}+\rho_\text{m}+\rho_{de0
}= \rho_{\text{r}0}s^{-4}+\rho_{\text{m}0} s^{-3}+\rho_{\rm de0}s^{-3(1+w_{\rm de0})}.
 \end{equation}

Now, Einstein's gravitational equation \eqref{fieldeqn} for  the FLRW metric \eqref{matrix} leads to the differential equations: 
\begin{equation}
 \label{FLRW1}
3\frac{\dot{s}^2}{s^2} = 8\pi\text{G}\rho - 3\frac{\kappa}{s^2},
 \end{equation}
\begin{equation}
 \label{FLRW2}
-2\frac{\ddot{s}}{s} - 3\frac{\dot{s}^2}{s^2}  = 8\pi\text{G}p - \frac{\kappa}{s^2}.
 \end{equation}

In this work, we discuss on the late-time universe. Without loss of generality, we consider  $8\pi G = 1 $, assume a vanishing radiation density parameter $\rho_r = 0$, and flat curvature ($\kappa = 0$), under which equations \eqref{FLRW1} and \eqref{FLRW2} reduce to :
\begin{equation}
 \label{EXFLRW1}
3\frac{\dot{s}^2}{s^2} = \rho,
 \end{equation}
\begin{equation}
 \label{EXFLRW2}
-2\frac{\ddot{s}}{s} - 3\frac{\dot{s}^2}{s^2}  = p_{\Lambda} .
 \end{equation}

The present equation of state (EoS) of vacuum dark energy ($w_{\Lambda0}=-1$) with time-varying cosmological constant, then, $-\rho_{\Lambda} = p_{\Lambda} = -{\Lambda}(t)$. Based on this assumption, the space-space fredimean equation in term of Hubble parameter $(H(t)= \dfrac{\dot{s}}{s})$ can be written as:

\begin{equation}
 \label{eq12}
2\dot H + 3H^2  = {\Lambda}.
 \end{equation}

Eq. \eqref{eq12}, in term of redshift with help of $\dfrac{d}{dt} = -H(z)(1+z)\dfrac{d}{dz}$, is given by

\begin{equation}
\label{eq13}
\frac{dH(z)}{dz} = \frac{3H(z)}{2(1+z)} - \frac{\Lambda(z)}{2H(z)(1+z)} .
\end{equation}

We adopt a cosmic time-varying cosmological constant \cite{ref50} defined  by the Hubble parameter as a function, read as:

\begin{equation}
\label{eq14}
\Lambda(t) = \alpha(\dot{H}+H^2) + \lambda H^2 + 4\pi G \rho \eta ,
\end{equation}
where $\alpha, \lambda, \eta $ are arbitrary constants.
The reformulation of the model is not an arbitrary choice but is strongly
motivated by both theoretical and observational grounds. On the theoretical side, terms
involving $H^2$ and $\dot{H}$ naturally emerge in renormalization-group approaches to
vacuum energy, quantum field theory in curved spacetime, and modified gravity frameworks,
thereby providing a physically meaningful basis for the adopted
parameterization. The inclusion of a density parameter-dependent contribution provides a phenomenologically consistent way to incorporate possible interactions between dark energy and matter. On the observational side, this parameterization offers clear phenomenological benefits: it significantly alleviates the Hubble tension (reducing it to 1.5$\sigma$),
 provides a better fit to DESI BAO, OHD, and PP\&SH0ES data, and captures quintessence-like behavior through the negative slope of the $Om$ diagnostic features unattainable within standard $\Lambda$CDM. Thus, the reformulated model is both theoretically well-motivated and empirically advantageous, offering a compelling alternative to address key shortcomings of $\Lambda$CDM

Now, rewriting the cosmological constant in terms of the redshift $z$ and using Eqs. \eqref{EXFLRW1}, and  \eqref{eq14} can be expressed as:

\begin{equation}
\Lambda(z) = \alpha \left[ -H(z)(1+z) \frac{dH}{dz} + H(z)^2 \right] + \lambda H(z)^2 + \dfrac{3}{2} H(z)^2 \eta.
\label{eq15}
\end{equation}

From Eqs. \eqref{eq13} and \eqref{eq15}, we have

\begin{equation}
(2-\alpha)\dfrac{dH}{H} = \dfrac{dz}{(1+z)}\left[3-\alpha- \lambda - \dfrac{3\eta}{2} \right] .
 \label{eq16}
\end{equation}

 After the integration of Eq. \eqref{eq16} , we obtain the governing equation for the Hubble parameter as,

 \begin{equation}
H(z) = H_0 (1+z)^{ \dfrac{3 - \alpha - \lambda - \dfrac{3\eta}{2}}{2 - \alpha} }\;,
\label{eq17}
\end{equation}
where $H_0$ is the present value of Hubble constant at $z = 0$.\\

Now, we calculate the vacuum energy density parameter $\Omega_{\Lambda } =  \dfrac{\Lambda}{3H^2} $ as:
\[
\Omega_\Lambda(z) = \frac{\alpha}{3} (1 - \gamma) + \frac{\lambda}{3} + \frac{\eta}{2},
\quad \text{where} \quad 
\gamma = \frac{3 - \alpha - \lambda - \frac{3\eta}{2}}{2 - \alpha}.
\]

%%----------------- Section -------------------------------------------

\section{DATA AND METHODOLOGY}

\begin{itemize}
\item \textbf{Dark Energy Spectroscopic Instrument (DESI BAO)}: These galaxy surveys cover a redshift range of $z\in[0.1,4.2]$ obtained from DESI BAO observations, which include the comoving spherically averaged distance into NB=7 distinct redshift bins. As noted in Ref. \cite{ref64}, the measurements across these bins are effectively independent; therefore, no covariance matrix is applied. Moreover, the associated systematic uncertainties typically introduce only a negligible offset \cite{ref64,ref65}. We employ refined distance measurements obtained from DESI BAO observations, which include the comoving spherically averaged distance $D_V(z)$, comoving luminosity distance $D_{M}(z)$, and Hubble distance $D_H(z)$ \cite{ref66,ref67}. These quantities are defined respectively as:
\begin{equation}
 D_V(z) \equiv \left[z D^2_M(z) D_H(z)\right]^{1/3},
\end{equation}

\begin{equation}
  D_{M}(z)=  \int_0^z \text{d}z' {c \over H(z')}, 
 \end{equation}
 and
 \begin{equation}
 D_H(z) = \frac{c}{H(z)},
 \end{equation}
where, $r_{\rm d}=\int_{z_{\rm d}}^\infty \frac{c_{\rm s}\text{d}z}{H(z)}$ is sound horizon at the drag redshift ($z_{\rm d}$) and ($c_{\rm s}$) is sound speed of the baryon–photon fluid. The corresponding comoving distances for various effective redshifts are summarized in in Table. \ref{tab1}.

\begin{table}[ht!]

\caption{\rm The statistics measurement of DESI BAO samples used in ref \cite{ref64}}.
\centering
\begin{tabular}{l|c|c|c|c}

\hline
\textbf{tracer} & $\bm{\,\,z_{\rm eff}\,\,}$  &  $\bm{\,\,D_{\rm V}(z)/r_{\rm d}\,\,}$ & $\bm{\,\,D_{\rm M}(z)/r_{\rm d}\,\,}$ & $\bm{\,\,D_{\rm H}(z)/r_{\rm d}\,\,}$  \\
\hline

\hline
BGS & $0.30$ & $7.93 \pm 0.15$ & --- & --- \\

LRG & $0.51$ & --- & $13.62 \pm 0.25$ & $20.98 \pm 0.61$ \\

LRG & $0.71$ & --- & $16.85 \pm 0.32$ & $20.08 \pm 0.60$ \\

 LRG + ELG & $0.93$ & --- & $21.71 \pm 0.28$ & $17.88 \pm 0.35$ \\

ELG & $1.32$ & --- & $27.79\pm0.69$ &$13.82\pm0.42$\\

QSO & $1.49$ & $26.07\pm0.67$ & -- & --- \\

Lya QSO & $2.33$ & --- &$39.71\pm0.94$ & $8.52 \pm 0.17$ \\
\hline
\hline
\end{tabular} 
\label{tab1}
\end{table}

\item\textbf{Observatinal Hubble Data (OHD)}: In this study, we employ a compilation of 33 measurements of the Hubble parameter \( H(z) \), obtained from observational Hubble data (OHD), spanning the redshift interval \( z \in (0.07, 1.965) \) \cite{ref67a,ref67b,ref67c,ref67d,ref67e,ref67f,ref67g,ref67h}. The foundational approach, initially formulated in Ref.~\cite{ref68}, relates the Hubble parameter to redshift and cosmic time through the relation:
\begin{equation}
H(z) = -\frac{1}{1+z} \frac{dz}{dt}. \tag{20}
\end{equation}

\item \textbf{PantheonPlus\&SH0ES (PP\&SH0ES)}:
We integrate the latest SH0ES Cepheid host distance calibrations \cite{ref69} into the likelihood function by incorporating distance modulus data from Type Ia supernovae (SNe Ia) within the Pantheon+ compilation \cite{ref70}. The PantheonPlus dataset consists of 1701 light curves corresponding to 1550 distinct SNe Ia events, spanning a redshift range of \( z \in [0.001, 2.26] \).\\

In our analysis, we utilize the Markov Chain Monte Carlo (MCMC) method to constrain cosmological parameters by analyzing astrophysical observational data, primarily focused on constrain the free parameter space $(H_0,l, \lambda, \eta )$ with corresponding the ranges $H_0 \in [60, 80]$, $l \in [-0.5, 0.5]$, $\lambda \in [0, 2]$, and $\eta \in [0, 1]$ respectively. The \texttt{emcee} library \cite{ref71} is employed for parallelized MCMC sampling using 100 walkers and 100000 steps to ensure convergence. By independently and jointly analyzing the 7 measurements from the DESI BAO survey, 33 observational $H(z)$ data points, and 1701 supernova measurements from the Pantheon+ and SH0ES compilations, we can extract meaningful constraints on the cosmological parameters and better understand the expansion history of the universe.\\

To statistically evaluate the consistency between the theoretical model and the observational data, we quantify the goodness-of-fit of our model by defining a statistical $\chi^2$ function based on the joint analysis of DESI BAO, OHD, and PP\&SH0ES data,

\begin{equation}
\chi^2_{\text{joint}} = \chi^2_{\text{DESI BAO}} + \chi^2_{\text{OHD}} + \chi^2_{\text{PP\&SH0ES}},
\end{equation}
where,

\begin{equation}
\chi^2_{\text{DESI BAO}} = \sum_{i=1}^{7} \frac{\left[ d^{obs}(z_i) - d^{th}(z_i) \right]^2}{\sigma^2_{d^{obs}(z_i)}},
\end{equation}

\begin{equation}
\chi^2_{\text{OHD}} = \sum_{i=1}^{33} \frac{\left[ H^{\text{obs}}(z_i) - H^{\text{th}}(z_i) \right]^2}{\sigma^2_{H^{\text{obs}}(z_i)}},
\end{equation}

\begin{equation}
\chi^2_{\text{PP\&SH0ES}} = \sum_{i,j}^{1701} \Delta\mu_i \left(C_{\text{PP\&SH0ES}}^{-1}\right)_{ij} \Delta\mu_j.
\end{equation}

In this analysis, $ d^{obs}(z_i)= D_{\rm V}(z_i)/r_{\rm d}$ and $ d^{th}(z_i)= D_{\rm V}(z_i)/r_{\rm d}$ denote the observed and model-predicted values of the DESI BAO distance ratio at redshift $z_i$, respectively, and $\sigma_{{d^{obs}(z_i)}}$ represents the associated observational uncertainty. Similarly, $H^{\text{obs}}(z_i)$ and $H^{\text{th}}(z_i)$ denote the observed and model-predicted values of the Hubble parameter at redshift $z_i$, respectively, and $\sigma_{H^{\text{obs}}(z_i)}$ represents the associated observational uncertainty. $\Delta \mu_i = \mu^{\text{th}}_i - \mu^{\text{obs}}_i$ represents the deviation between the theoretical and observed distance modulus values. The matrix $C_{\text{PP\&SH0ES}}^{-1}$ is the inverse of the covariance matrix corresponding to the Pantheon dataset, which accounts for statistical correlations between supernova measurements.
\end{itemize}

\section{Results and discussion}

\begin{figure*}[hbt!]
    \centering
    \includegraphics[width=0.8\linewidth]{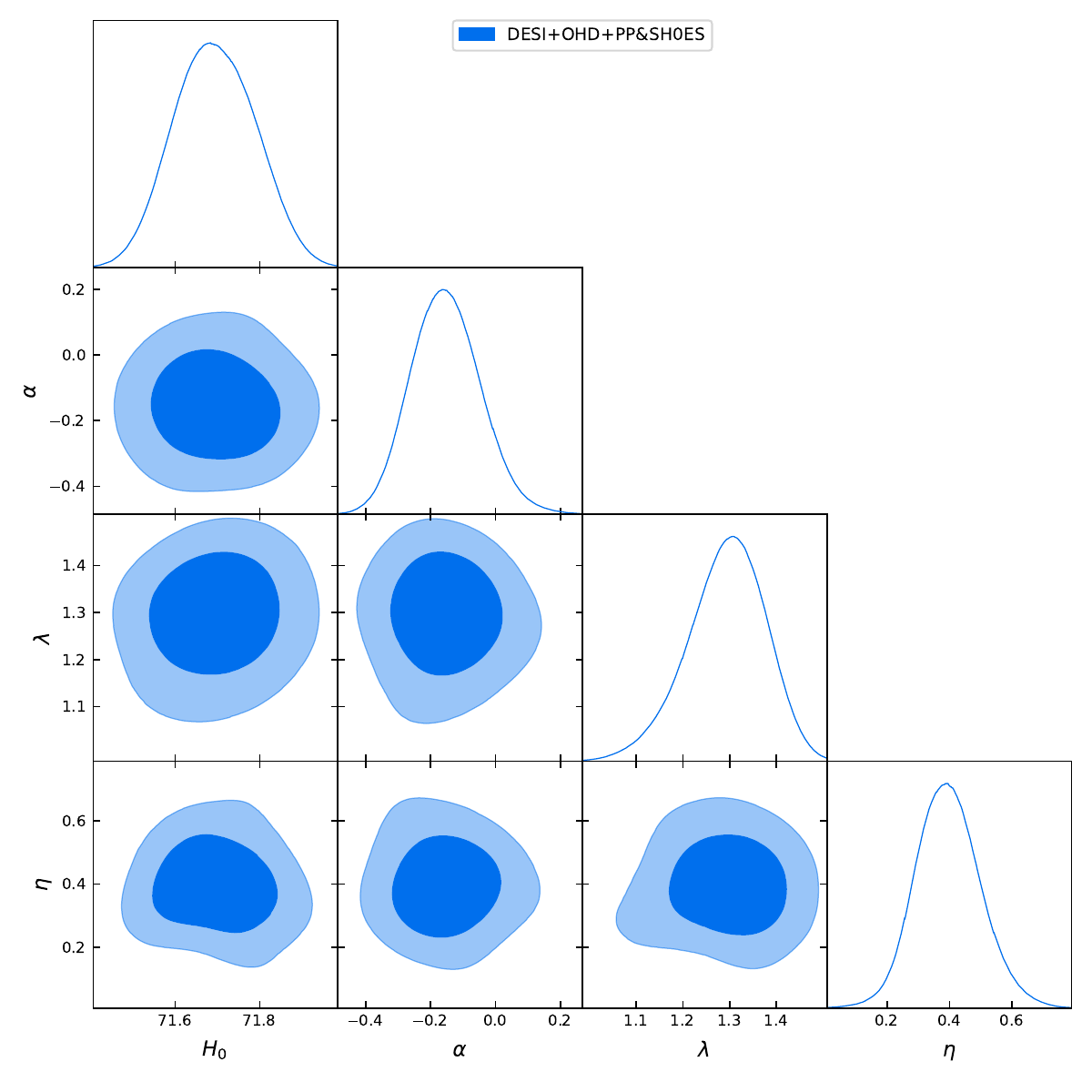}
    \caption{The free parameters constraints of the $\Lambda(t)$CDM model are presented at the 1$\sigma$ and 2$\sigma$ confidence levels, based on the combined analysis of DESI BAO, OHD, and PP\&SH0ES data.}

    \label{fig1}
\end{figure*}

In this paper, we investigate the free parameters $(H_0, \alpha, \lambda, \eta)$ along with the derived parameters of the Power law model using observational datasets such as DESI BAO, OHD, and PP\&SH0ES. The primary focus is to examine the influence of the additional free parameters $\alpha, \lambda, \eta$ on other key cosmological parameters. Fig.\ref{fig1} presents the triangle plot showing the one- and two-dimensional marginalized distributions of the parameters $H_0, l, \lambda, \eta$ at both 68\% as well as  95\% confidence levels (with the dark blue regions representing 68\% C.L. and the light blue regions representing 95\% C.L.). In this figure clearly demonstrates smooth distributions and well-constrained bounds for all four parameters under the combined datasets.\\

\begin{figure*}[hbt!]
    \centering
    \includegraphics[width=0.9\linewidth]{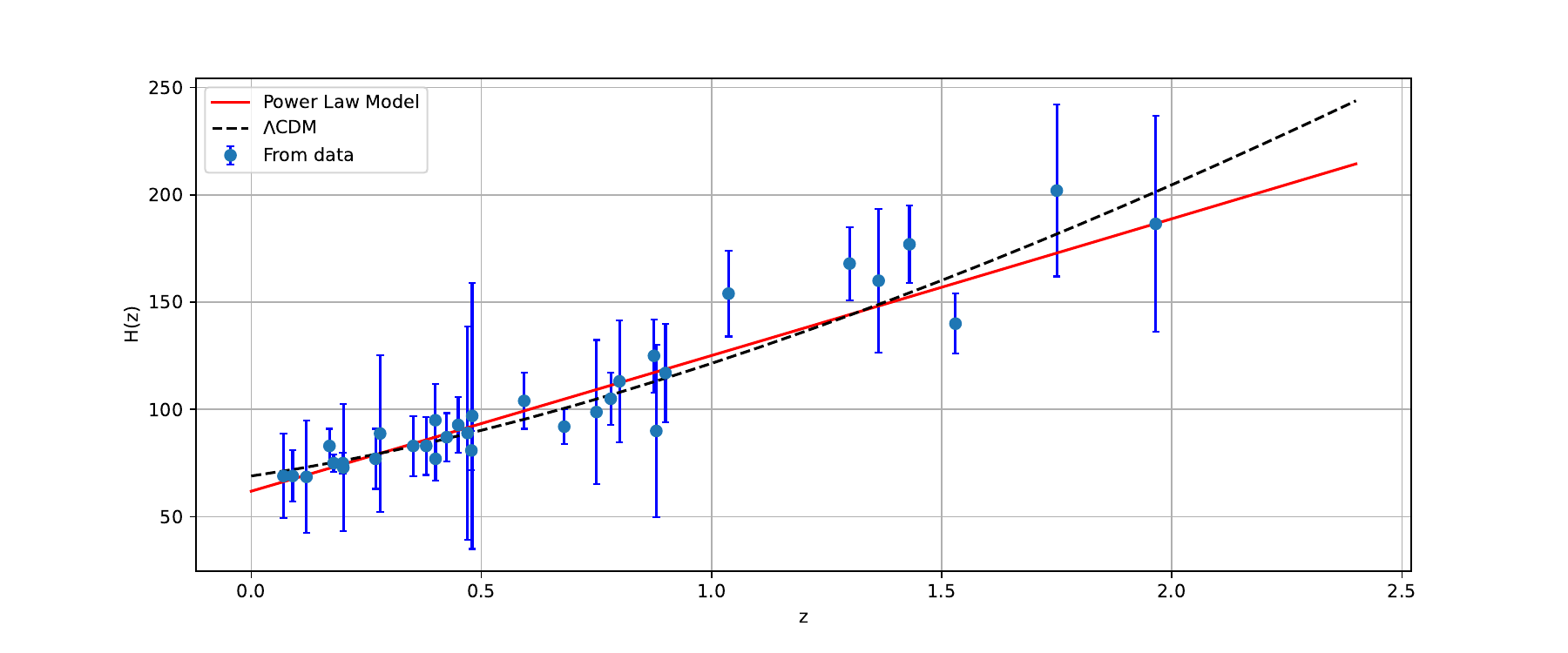}
    \caption{The 2D plot shows how the Power Law model diverges/converges from the $\Lambda$CDM model, in comparison with 33  observation Hubble parameter data points with corresponding error bars.
}
    \label{fig2}
\end{figure*}

In this context, the Hubble constant $H_0$ is one of the most fundamental parameters in cosmology, representing the current expansion rate of the universe. It plays a central role in determining the age, size, and evolution of the universe. However, despite significant advancements in observational techniques, a notable discrepancy has emerged between different methods of measuring $H_0$, leading to what is commonly referred to as the Hubble tension. This tension arises from the inconsistency between early-universe measurements—primarily those inferred from the CMB by the Planck satellite under the assumption of the $\Lambda$CDM model—and late-universe measurements based on the local distance ladder, such as those from the SH0ES collaboration using supernovae calibrated by Cepheid variables, are approximately $5\sigma$. In the present work, we investigate the value of $H_0$ within the framework of the Power law model $\Lambda$(t)CDM using recently available observational datasets, including OHD, PP\&SH0ES, and DESI BAO. As summarized in Table \ref{tab2}, Our analysis yields a constraint of \(H_0 = 71.9 \pm 0.23\) km s\(^{-1}\) Mpc\(^{-1}\) when employing the PP\&SH0ES dataset independently, and a similar value is corroborated by the joint analysis incorporating  DESI BAO, OHD, and PP\&SH0ES datasets. Here, we compare our findings  with SH0ES collaboration's measurement of \(H_0 = 73.27 \pm 1.04\) km s\(^{-1}\) Mpc\(^{-1}\)  at the 68\% CL., based on supernovae calibrated with Cepheid variables \cite{ref72}. We find that $1.30\sigma (1.50\sigma)$ tensions arise in the time-varying dark energy model $\Lambda$(t)CDM with PP\&SH0ES(DESI BAO+OHD+PP\&SH0ES) datasets, respectively. This Hubble tension is reduced $\Lambda$(t)CDM compared to the $\Lambda$CDM model, which is statistically significant; we conclude that our model is consistent with the SH0ES calibration and supports a higher value of the Hubble constant. However, the discrepancy with the Planck collaboration's determination of \(H_0 = 67.27 \pm 0.60\) km s\(^{-1}\) Mpc\(^{-1}\),\cite{ref73} also at a 68\% confidence level, persists as significant across all datasets evaluated. In Fig. \ref{fig2}, we observe the observational and theoretical values of the Hubble parameter relative to redshift over the range $z \in [0,1.96]$. Our model shows moderate agreement with the observational Hubble data.\\

\begin{figure}[hbt!]
    \centering
    \includegraphics[width=0.9\linewidth]{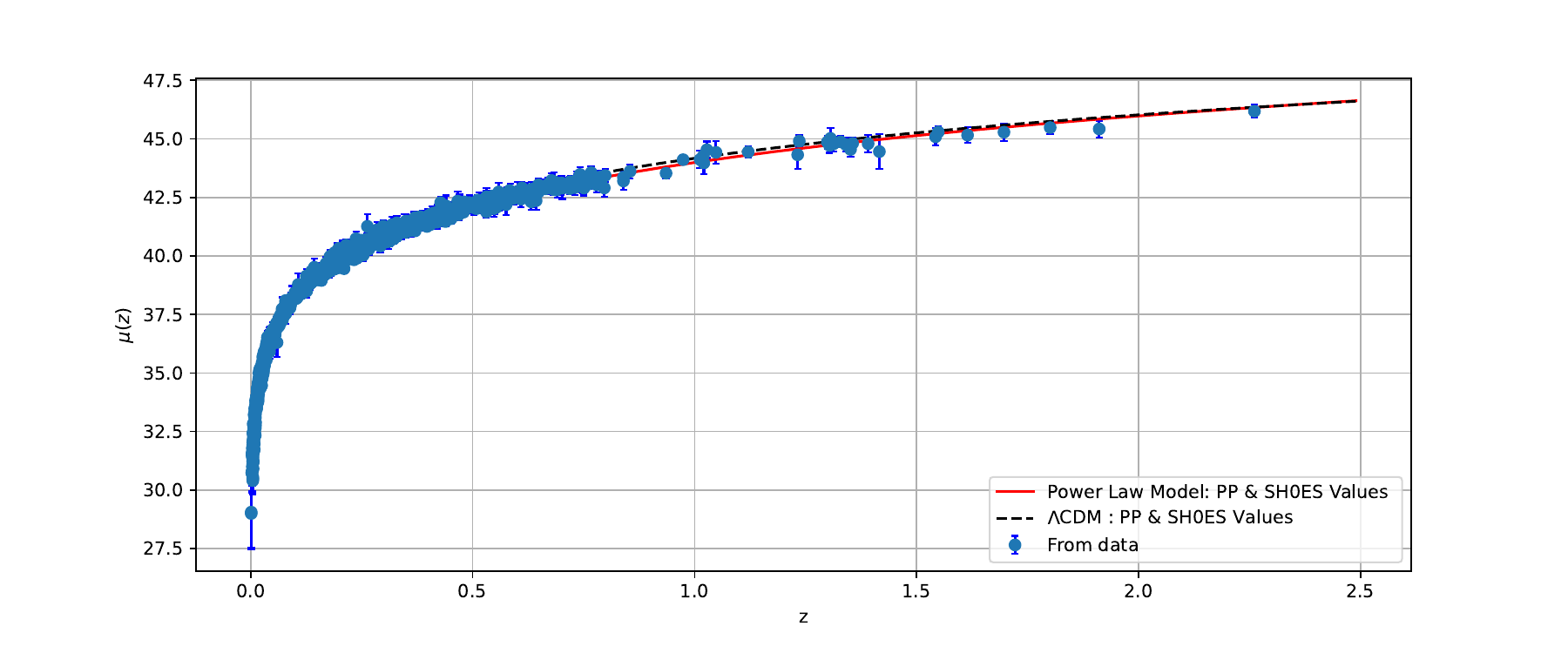}
    \caption{The 2D plot shows the distance
modulus $\mu(z) $ for the Power law model (red line) and the $ \Lambda$CDM model (black dotted line), alongside 1701  Pantheon Plus and SH0ES data points with corresponding blue colour error bars.}
    \label{fig3}
\end{figure}

Next, we compare our theoretical estimation of the distance modulus $\mu$ with $1701$ observational data points from the Pantheon Plus \& SH0ES compilation, based on the SH0ES calibration at redshift interval $z\in[0,2.2613]$. We find that our model is good agreement with SH0ES  observation datasets. To quantify closeness, of the theory and observation  numerical value of $\mu$ with the help of Chi-square, which is defined as 
$$
\chi^2 = \frac{(\mu_{\text{obs}} - \mu_{\text{th}})^2}{\mu_{\text{err}}^2},
$$
where $\mu_{\text{obs}}$ and $\mu_{\text{th}}$ represent the observed and theoretical distance moduli, respectively, and $\mu_{\text{err}}$ denotes the observational error associated with $\mu_{\text{obs}}$. The resulting chi-square value indicates that our model achieves a $96.73\%$ best-fit alignment with the observational data. This strong agreement is visually supported in Fig. \ref{fig3}, which illustrates the close correspondence between the theoretical curve and the observed supernova data.\\

$\bf{Om(z)}$ \textbf{diagnostic}: Also, the $Om(z)$ diagnostic is a significant parameter in cosmology that serves as a valuable geometric tool to test the $\Lambda$CDM model and distinguish it from alternative  (DE) models with different observational datasets. It relies solely on the Hubble parameter $H(z)$, making it both simple to apply and interpret, as it depends only on the first-order derivative of the scale factor. For a spatially flat universe (curvature is zero), the $Om(z)$ diagnostic provides a null test for the $\Lambda$CDM model with a constant slope, which indicates that the EoS of dark energy favors the cosmological constant ($\omega = -1$). Meanwhile, Deviations from $Om(z)$ diagnostic behavior offer insights into the nature of dark energy:  positive diagnostic slope suggests a Quintessence-like behavior ($\omega > -1$), while negative diagnostic slope points to a Phantom-like nature ($\omega < -1$) \cite{ref74,ref75}. Thus, $Om(z)$ functions as an effective and model-independent method to probe the dynamics of dark energy through observational data of the Hubble parameter, which defined as:

\begin{equation}
Om(z) = \frac{H^2(z)/H_0^2 - 1}{(1+z)^3 - 1}.
\end{equation} 

\begin{figure}[hbt!]
    \centering
    
    \includegraphics[width=0.45\linewidth]{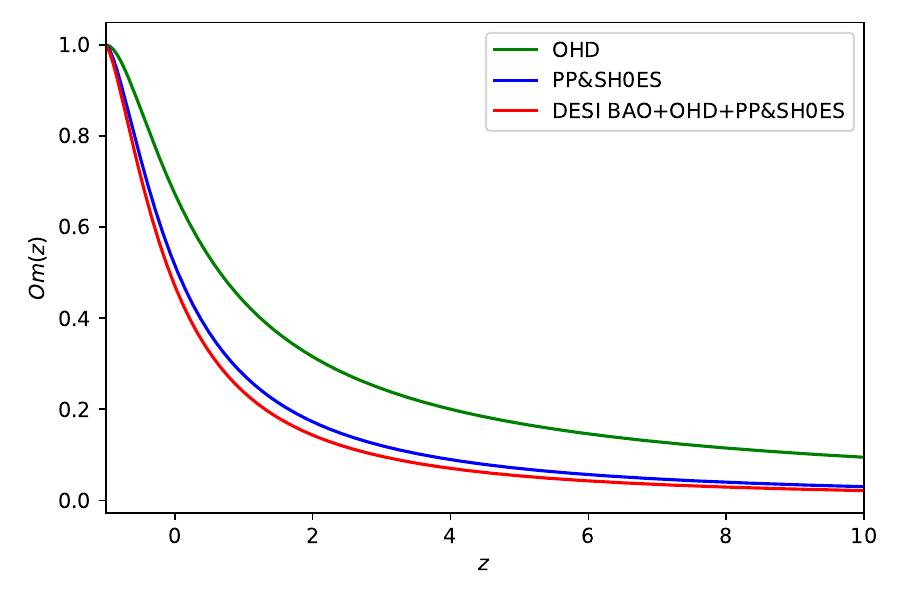}
    \includegraphics[width=0.47\linewidth]{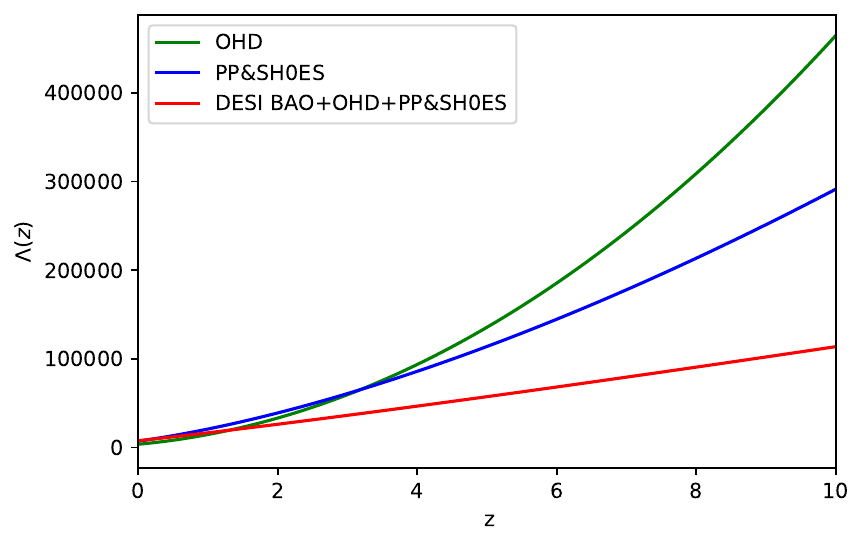}
    \caption{The left panel displays Om(z) diagnostic with respect to $z$ and the right panel shows $\Lambda$(z) with respect to $z$ with Power law model from all the considered datasets. }
    \label{fig4}
\end{figure}

We have discuss $Om(z)$ diagnostic within the framework of the Power Law model using different observational datasets, including OHD, PP\&SH0ES, and joint analysis DESI BAO + OHD + PP\&SH0ES. As illustrated in the left panel of Fig.\ref{fig4}, the $Om(z)$ diagnostic has a consistently negative slope with all considered data sets. This behavior indicates that the power law model under investigation tends to favor a  Quintessence-like nature of dark energy, characterized by the EoS parameter $\omega >-1$,  thereby deviating from the standard $\Lambda$CDM model.The right panel of the figure presents the redshift evolution of the dynamic cosmological constant $\Lambda$(z) within the framework of a time-dependent $\Lambda$CDM model. Three distinct observational datasets are compared. The OHD data showing the most pronounced growth in $\Lambda$(z), which increases steeply. When we incorporates the Pantheon+ supernova sample along with SH0ES $H_{0}$ measurements. This datasets predict a moderately slower rise in $\Lambda(z)$.Combines multiple cosmological probes, including DESI BAO data, resulting in the lowest growth rate of $\Lambda$(z). The curve remains significantly suppressed, indicating that multi-probe constraints lead to tighter bounds on the evolution of the cosmological constant $\Lambda(z)$.
Overall, the plot illustrates that including additional datasets (particularly large-scale structure and BAO measurements) systematically reduces the growth rate of $\Lambda$(z) with increasing redshift $z$, reflecting stronger constraints on deviations from a strictly constant $\Lambda$. This behavior supports the idea that combined observational analyses favor a cosmological constant that evolves more mildly with cosmic time.\\

Here, we examine the boundaries and consistency of the energy density parameter \(\Omega_{\Lambda 0}\) within the context of our model. As summarized in Table \ref{tab2}, the numerical mean value of \(\Omega_{\Lambda 0}\) is found to be $0.62\pm 0.055$ from the joint analysis and $0.48 \pm 0.033$ when using the PP\&SH0ES dataset alone. Our result for \(\Omega_{\Lambda 0}\) is in agreement with the initial findings from the MAXIMA flight combined with COBE-DMR, which gave the range $0.45 < \Omega_{\Lambda 0} < 0.75$. Furthermore, observations of Type Ia supernovae combined with constraints from the CMB, gravitational lensing, and stellar dynamical analysis suggest a value of $\Omega_{\Lambda0} \sim 0.7$. Sievers et al. \cite{ref76} and Spergel et al. \cite{ref77} reported values of \(\Omega_{\Lambda0} = 0.67 \pm 0.10\) and \(0.719 \pm 0.029\), respectively. Considering various observational datasets, many researchers prefer a range of \(0.40 \leq \Omega_{\Lambda0} \leq 0.785\). In the right panel of Fig.\ref{fig4} illustrates the behavior of the time-varying cosmological constant with respect to  $z$, showing that it is an increasing function of $z$, which is consistent with observational evidence.

\begin{table*}[hbt!]
     \caption{Marginalized constraints, mean values with 68\% CL, on the free and some derived parameters of the $\Lambda(t)$CDM models for different dataset combinations.}
     \label{tab2}
     \scalebox{0.85}{
 \begin{tabular}{lccc}
  	\hline
    \toprule
   \textbf{Dataset }&\;\textbf{OHD}\;&\; \textbf{PP\&SH0ES}\;& \;\;\;\;\;\textbf{DESI BAO+OHD+PP\&SHOES}
   \\ \hline
   
\vspace{0.4cm}
{\boldmath${H_0}$}&$61.9^{+0.150}_{-0.140}   $ &$71.9\pm 0.23$ &$71.70\pm 0.15$\\
\vspace{0.4cm}
{\boldmath$\alpha$}&$0.02 \pm 0.11   $ &$0.01\pm 0.10$ &$-0.148^{+0.09}_{-0.11}$\\
\vspace{0.4cm}
{\boldmath$\lambda$}&$0.863\pm0.089   $ &$1.33\pm ^{+0.10}_{-0.09}$ &$1.33\pm 0.10$\\
\vspace{0.4cm}
{\boldmath$\eta$}&$0.070^{+0.14}_{-0.12} $ &$0.084\pm 0.0091$ &$0.402\pm 0.093$\\
\vspace{0.4cm}

\vspace{0.4cm}
{\boldmath$\Omega{}_{\Lambda}$}&$0.32\pm 0.072   $ &$0.48 \pm 0.033$ &$0.62 \pm 0.055$\\

\hline
 \hline
\end{tabular}
}
\end{table*}
\textbf{EoS Parameter:} The total (EoS) parameter is the ratio of total isotropic fluid pressure to the total energy density of the universe, which characterizes different phases of cosmic evolution. When $\omega_{\text{tot}} = 0$, that means the universe is in a dust-dominated phase, indicative of non-relativistic matter prevalence. while $\omega_{\text{tot}} = \frac{1}{3}$ corresponds to a radiation-dominated era. At the other end of the spectrum, $\omega_{\text{tot}} = -1$ represents vacuum energy domination as described in the $\Lambda$CDM model, and indicates the current epoch of accelerated expansion in the universe. Recent cosmological studies has emphasized the accelerating phase of the universe, identified by $\omega_{\text{tot}} < -\frac{1}{3}$, This regime includes the quintessence regime ($-1 < \omega_{\text{tot}} < -\frac{1}{3}$) as well as the phantom regime ($\omega_{\text{tot}} < -1$). These scenarios describe an increasingly rapid universe expansion, with the phantom regime indicating an even more extreme acceleration.The (EoS) parameter is defined as $\omega_{tot} = p_{eff}/\rho_{eff}$. We obtain the present values of $\omega_{\text{tot}}$  from the power-law model using OHD, PP\&SH0ES, and DESI+OHD+PP\&SH0ES datasets, which are approximately -0.32, -0.48, and -0.66, respectively. These results indicate that $\omega_{\text{tot}}$ lies within the quintessence regime $-1 < \omega_{\text{tot}} < -\frac{1}{3}$ and dark energy domination of the current era in the universe. \\ 

\begin{figure}[hbt!]
    \centering
    \includegraphics[width=0.6\linewidth]{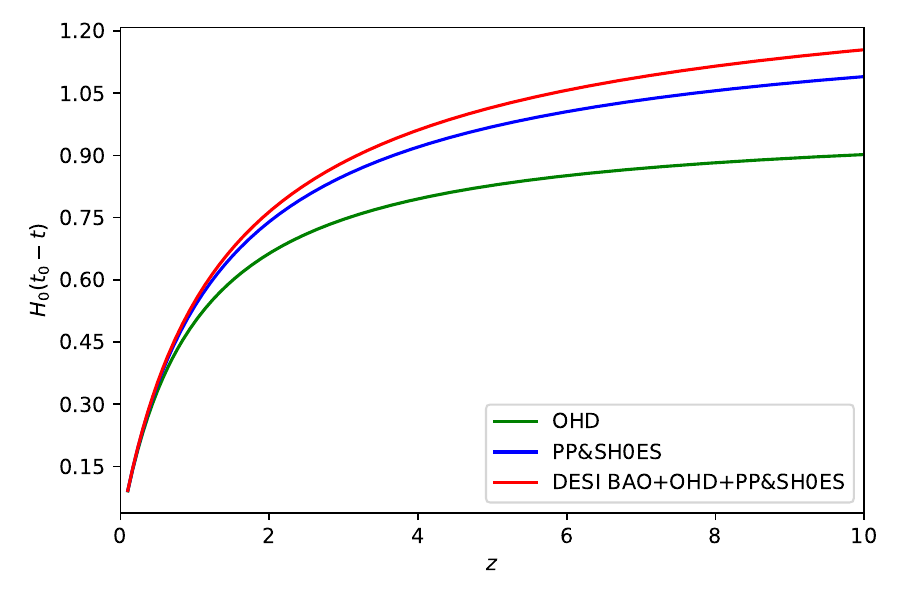}
    \caption{The 1D plot $H_0(t_0 - t)$ vs z with Power law model from OHD, PP\&SH0ES, and DESI+OHD+PP\&SH0ES datasets }
    \label{fig5}
\end{figure}

\textbf{Age of the Universe:} The age of the universe can be estimated using the lookback time, which represents the time interval between the present age of the universe, $t_0$, and its age at a given redshift $z$. In a cosmological model where the Hubble parameter $H(z)$ varies with redshift, the lookback time is calculated through the following integral.

We calculate the age of the universe given by 
\begin{equation}
\label{eq26}
t_0 - t(z) = \int_0^z \frac{dz'}{(1 + z') H(z')}
\end{equation}

from \eqref{eq17} and \eqref{eq26}, we get

\[
H_0 (t_0 - t(z)) = \int_0^z (1 + z')^{\frac{2 - \alpha}{5 - 2\alpha - \lambda - \frac{2}{3}\eta}}\, dz'
\].

 The present age of the universe, label as $t_0$ is given by

 \[
H_0 t_0 = \int_0^{\infty} (1 + z')^{\frac{2 - \alpha}{5 - 2\alpha - \lambda - \frac{2}{3}\eta}}\, dz'
\]
Fig.\ref{fig5} illustrates the variation of the normalized lookback time, denoted as $H_{0}(t_{0}-t)$, as a function of redshift 
$z$, for three different observational datasets: OHD, PP\&SH0ES (blue curve), and the combined dataset DESI BAO+OHD+PP\&SH0ES (red curve). The lookback time $(t_{0}-t)$ represents the temporal separation between the present epoch and the cosmic time corresponding to a given redshift $z$, and is normalized here by the Hubble constant $H_{0}$ to render it dimensionless and suitable for comparative analysis. All three curves exhibit a monotonically increasing trend with redshift, consistent with cosmological expectations, wherein higher redshifts correspond to earlier epochs in the universe's history. It is important to note
that the empirical value of age of the Universe in Planck
collaboration results \cite{ref78} is obtained as $t_{0} =13.81 \pm 0.038$
Gyrs. In some other cosmological investigations, age of
the Universe is estimated as $14.46\pm 0.8$ Gyrs \cite{ref79}, $14.3\pm
0.6$ Gyrs \cite{ref80}, and $14.5 \pm 1.5$ Gyrs \cite{ref81}. In this paper, the present age of universe for the derived model is estimated as $t_{0} =13.74\pm1.5$ for OHD , $t_{0} =14.88\pm1.7$ PP$\&$ SHOES and $t_{0} =15.62\pm0.6$ for DESI BAO+OHD+PP$\&$ SHOES.\\

\textbf{Statistical Analysis}: In the final step of the observational analysis, we are comparing the power-law and standard models through standard information criteria such as  Akaike Information Criterion (AIC) \cite{ref81a} and the Bayesian, or Schwarz, Information Criterion (BIC) \cite{ref81b}. Both of these are widely employed in statistics and data analysis, which are described below.
$$\text{AIC}  = -2 \ln \mathcal{L}+ 2 d = \chi^2_{min} + 2 d,$$

$$\text{BIC}  = -2 \ln \mathcal{L}+  d \ln N = \chi^2_{min} +  d \ln N,$$

In this expression, the maximum likelihood is given by $\mathcal{L} = \exp\left(-\chi_{\text{min}}^2/2\right)$, with $d$ representing the number of free parameters in the model and $N$ the total number of data points included in the analysis. To evaluate and compare different models, it is essential to adopt a baseline scenario, which is the standard $\Lambda$CDM cosmology. For any given model $M$, one can assess its relative performance by calculating the difference $\Delta X = X_M - X_{\Lambda\text{CDM}}$, where $X$ corresponds to either the AIC or BIC criterion. Based on the value of $\Delta X$, the model's level of support can be interpreted as follows: For the AIC, Strong support if $\Delta \mathrm{AIC} \leq 2$ (indicating a good fit), Moderate support if $4 \leq \Delta \mathrm{AIC} \leq 7$, and Essentially no support if $\Delta \mathrm{AIC} \geq 10$. For the BIC, the evidence is considered positive if \( 2 \leq \Delta \mathrm{BIC} \leq 6 \), strong if \( 6 < \Delta \mathrm{BIC} \leq 10 \), and \emph{very strong} if \( \Delta \mathrm{BIC} > 10 \). \\ 

\begin{table*}[hbt!]
	\caption{The difference, $\Delta \text{AIC} = \text{AIC}_{\text{our model}} - \text{AIC}_{\Lambda \text{CDM}}$ and $\Delta \text{BIC} = \text{BIC}_{\text{our model}} - \text{BIC}_{\Lambda \text{CDM}}$ for our model with respect to $\Lambda$CDM  from all considered data sets.}
	\label{tab3}
	\scalebox{0.85}{
		\begin{tabular}{lccc}
			\hline
			\toprule
			\textbf{Dataset }&\;\;\;\;\;\;\textbf{OHD}\;\;\;\;\;\;&\;\;\;\;\;\;\ \textbf{PP\&SH0ES}\;& \;\;\;\;\;\textbf{DESI+OHD+PP\&SH0ES}
			\\ \hline
			\textbf{Model} & \textbf{\text{Our model}}\,&\textbf{\text{Our model}}\,&\textbf{\text{Our model}}\\\
			&\textcolor{teal}{\textbf{$\bm{\Lambda}$CDM}}\, & \textcolor{teal}{\textbf{$\bm{\Lambda}$CDM}}\, & \textcolor{teal}{\textbf{$\bm{\Lambda}$CDM}} 
			\\ \hline

			\vspace{0.1cm}
			{\boldmath$\rm AIC$}&$ 24.30$ &$ 1298.10$ &$1340.42 $\\
			
			&\textcolor{teal}{$20.62$} &\textcolor{teal}{$1292.14$} &\textcolor{teal}{$1334.66$} \\
			\hline
			\vspace{0.1cm}
			{{\boldmath$\rm \Delta AIC$}}&$ 3.68$ &$ 5.96$ &$ 5.76  $\\

			\hline
			\vspace{0.1cm}
			{{\boldmath$\rm BIC$}}&$ 30.28 $&$1319.85$&$1362.26 $  \\
			
			& \textcolor{teal}{$ 25.10$}& \textcolor{teal}{$ 1308.45$} &\textcolor{teal}{$ 1351.04$}\\
			\hline
			\vspace{0.1cm}
			{{\boldmath$\rm \Delta BIC$}}&$5.17 $& $11.39 $&$11.21 $ \\
			
			\hline
			\hline

			\vspace{0.1cm}
			{{\boldmath$\chi^2_{min} $}}&$16.30 $& $1290.10 $&$1332.42 $ \\
			
			& \textcolor{teal}{$ 14.62$}& \textcolor{teal}{$ 1286.14$} &\textcolor{teal}{$ 1328.66$}\\
			
			\hline
			\hline
		\end{tabular}
	}
\end{table*}

Table \ref{tab3} presents the computed values of AIC, BIC, and $\chi^2$ for our models, compared against the standard $\Lambda$CDM cosmology, using three different observational datasets: OHD, PP\&SH0ES, and the combined DESI BAO+OHD+PP\&SH0ES compilation. To ensure a fair comparison, we also report the relative differences $\Delta$AIC and $\Delta$BIC with respect to the baseline $\Lambda$CDM model. It is important to note that our model's AIC and BIC values are slightly larger than those of the $\Lambda$CDM scenario from all the given datasets. These higher values arise from our model due to extra degrees of freedom. Several papers are presented in the literature which have higher AIC and BIC values relative to the standard model \cite{ref82,ref83,ref84},  where cosmological models-despite sometimes yielding competitive or even superior fits-often result in higher AIC and BIC values due to their enlarged parameter spaces. Hence, our results remain statistically consistent and in good agreement with the observational data.

\section{Conclusion}
In this study, we considered a model of cosmological constants varying over time, $\Lambda$(t)CDM, where the cosmological constant $\Lambda$(t) is defined as a function of the Hubble parameter and its derivative. The model includes three free parameters: $\alpha$, $\lambda$, and $\eta$. Using Markov Chain Monte Carlo (MCMC) analysis with the emcee algorithm, we constrained the model parameters $(H_{0}, \alpha, \lambda, \eta)$ using the combined observational datasets. DESI BAO, 33-point Hubble parameter measurements, and PP\&SH0ES. The best-fit $H_{0}$ value obtained from the PP\&SH0ES dataset alone is $71.9\pm 0.23$ km/s/Mpc, consistent with the value obtained from the joint analysis as shown in Fig.\ref{fig1}. This result is about $\approx 1.3\sigma(1.5\sigma)$ away from the SH0ES value $(73.27\pm1.04)$, indicating a significant reduction in the Hubble tension, although the tension with Planck $(67.27\pm 0.60)$ remains.\\

Fig.\ref{fig2} illustrates the evolution of the Hubble parameter $H(z)$ with redshift $z$, comparing observational data (with error bars) against two theoretical models: the Power Law Model (red solid line) and the $\Lambda$CDM model (black dashed line). The observational data consist of 33 discrete measurements of $H(z)$, derived from cosmic chronometers or similar cosmological probes. The error bars represent uncertainties in the observational determination of $H(z)$. The distance
modulus $\mu(z) $ for the power law model  and the $ \Lambda$CDM model is shown in Fig.\ref{fig3}. The reconstructed $Om(z)$ diagnostic consistently exhibits a negative slope, suggesting a quintessence-like dark energy behavior with $\omega>-1$, thus deviating from the constant dark energy density assumed in the standard $\Lambda$CDM model (see Fig.\ref{fig4}) left side. Further analysis reveals that the present-day value of the dark energy density parameter, $\Omega_{\Lambda0}$, estimated from our model is compatible with several previous observational results.\\ 

The present-day effective total equation of state derived from the best-fit parameters is approximately: $\omega_{\text{tot}} \approx -0.32$  \text{(OHD only)}, $\omega_{\text{tot}} \approx -0.48 $ \text{(PP\&SH0ES)}, $\omega_{\text{tot}} \approx -0.66 $ \text{(Joint data analysis)}. All values fall within the accelerating regime ($-1 < \omega_{\text{tot}} < -\frac{1}{3}$), confirming that the universe is currently undergoing dark energy-driven acceleration, with quintessence-like characteristics. The $\Lambda$(t)CDM model presented here provides a good fit to current observational data, particularly at late times, and offers a promising extension to the standard $\Lambda$CDM model. It partially alleviates the Hubble tension and allows for a time-evolving dark energy component. Future high-precision observations (from upcoming BAO surveys, gravitational waves, or improved CMB measurements) will be essential to further test and refine this model.Moreover, the evolution of the normalized lookback time 
$H_{0}(t_{0}-t)$ versus redshift $z$, as illustrated in Fig.5, indicates
that models constrained by combined datasets (DESI BAO+OHD+PP$\&$SH0ES) 
yield consistently higher lookback times compared to those using only OHD.
This reflects a comparatively slower late-time expansion rate and provides 
tighter constraints on the cosmic age–redshift relation under the time-dependent $\Lambda(t)$ framework.\\

Overall, the power-law $\Lambda(t)$CDM model presents itself as a viable and compelling alternative to the standard $\Lambda$CDM paradigm. It offers better compatibility with recent observational data, provides a natural framework to interpret the dynamics of dark energy, and significantly mitigates the Hubble tension. These results suggest that considering a time-varying cosmological constant can enrich our understanding of the universe's late-time acceleration and motivate further theoretical and observational investigations in modern cosmology.

\section*{Declaration of competing interest}
We wish to confirm that there are no known conflicts of interest associated with this publication and there has been no significant
financial support for this work that could have influenced its outcome.

\begin{acknowledgments}
\noindent 
 
\noindent 
 The authors (A. Dixit \& A. Pradhan) are thankful to IUCAA, Pune, India for providing support and facility under Visiting Associateship program. M.Yadav is supported by a Junior Research Fellowship (CSIR/UGC Ref.\ No.\ 180010603050) from the University Grants Commission, Govt. of India. The authors thank the Reviewer and the Editor for their constructive comments, which have enhanced the paper in its current form.   

\end{acknowledgments}

\section*{Data availability } No data was used for the research described in the article.

\end{document}